\documentclass[11pt,twoside,letterpaper]{article} 
\usepackage{times,fancyhdr}
\usepackage[dvips]{graphicx}
\makeatletter 
\def\cleardoublepage{\clearpage\if@twoside \ifodd\c@page\else%
    \hbox{}%
    \thispagestyle{empty}%
    \newpage%
    \if@twocolumn\hbox{}\newpage\fi\fi\fi} 
\makeatother 
\renewcommand{\thesection}{\arabic{section}.}
\renewcommand{\thesubsection}{\thesection\arabic{subsection}.}

\def\figurename{Figure}
\makeatletter
\renewcommand{\fnum@figure}[1]{\figurename~\thefigure.}
\makeatother
\def\tablename{Table}
\makeatletter
\renewcommand{\fnum@table}[1]{\tablename~\thetable.}
\makeatother
\begin{document}
\title{
{\begin{flushleft}
\vskip 0.45in
{\normalsize\bfseries\textit{Chapter~1}}
\end{flushleft}
\vskip 0.45in
%
%
%
%
\bfseries\scshape Terrestrial and Astrophysical Superfluidity: cold atoms and neutron matter}}
\author{\bfseries\itshape Alexandros Gezerlis\\
Department of Physics, University of Washington, \\Seattle, Washington 98195 USA\\
\\
\bfseries\itshape J. Carlson\\
Theoretical Division, Los Alamos National Laboratory, \\Los Alamos, New Mexico 87545 USA}
\date{}
\maketitle

\abstractname{: After a brief historical overview of superfluidity in connection
with neutron matter and cold fermionic atoms, we discuss the commonalities
between these two systems as well as their relevance to the physics of neutron
star crusts. We then present the methodological tools we use to attack the many-body
problem, comparing Quantum Monte Carlo to mean-field theory
and to the analytic expectations at weak coupling. We review recent results on
the equation of state and the pairing gap of cold atoms and neutron matter and contrast
them with a variety of calculations by other groups. We conclude by giving
a few directions of possible future inquiry.}

\thispagestyle{empty}
\setcounter{page}{1}
\thispagestyle{fancy}
\fancyhead{}
\fancyhead[L]{In: Neutron Star Crust \\ 
Editors: C.A. Bertulani and J. Piekarewicz, pp. {\thepage-\pageref{lastpage-01}}} 
\fancyhead[R]{ISBN 0000000000  \\
\copyright~2012 Nova Science Publishers, Inc.}
\fancyfoot{}
\renewcommand{\headrulewidth}{0pt}
\vspace{0.2in}

\pagestyle{fancy}
\fancyhead{}
\fancyhead[EC]{Alexandros Gezerlis}
\fancyhead[EL,OR]{\thepage}
\fancyhead[OC]{Terrestrial and Astrophysical Superfluidity}
\fancyfoot{}
\renewcommand\headrulewidth{0.5pt} 
%

\section{Historical Overview}
\label{section:INTRO_histo}

One could fill volumes (and many people have done so) writing on the history
of superfluidity and superconductivity. Our goals are much humbler: we only
wish to connect the pairing found in neutron matter with recent cold fermionic
atom experiments. Thus, our ever-shifting kaleidoscope
(which relies mainly on Refs.
\cite{Bardeen:1990,Shapiro:1983,Ketterle:2008,Giorgini:2008}) by construction
passes over in silence discoveries that are major, but do not bear a direct
relation to the physics of neutron-star crusts. The natural way to start such a
brief historical overview is by mentioning Heike Kamerlingh Onnes, who in 1908
liquefied $^4$He, at 4.2 K. Onnes mentions in his 1913 Nobel lecture
\cite{Onnes:1913} that the behavior of $^4$He at an even colder temperature,
2.2 K, presents an interesting problem, which ``could possibly be connected
with the quantum theory'' (note that this was said in 1913). Onnes had already
realized that there was something peculiar about that temperature. Instead of
staying on the subject of bosonic systems, he used $^4$He to cool down
mercury, discovering superconductivity (a drop of the resistivity to
non-measurable values) at 4.1 K in that system. 

Then, in 1925, when the statistics different particles follow was
still unresolved, Einstein predicted the phenomenon now known as Bose-Einstein
condensation (BEC), according to which below a certain temperature a finite
fraction of the total number of particles occupies the lowest-energy
single-particle state. In 1927, Keesom and Wolfke made the distinction between
the two phases of $^4$He above and below the $\lambda$ transition point around
2.2 K. In 1938 Kapitsa and, separately, Allen \& Misener discovered
superfluidity (a drop of the viscosity to essentially zero) using $^4$He. A
year later, Kapitsa managed to get Lev Davidovich Landau released from prison,
with the hope that he would be able to provide a theoretical
explanation of this new effect. 

In an apparently unrelated context, Baade and Zwicky (in 1934, only two
years after the discovery of the neutron) proposed the idea of high-density
and small-radius {\it neutron stars} which would be much more gravitationally
bound than ordinary stars and would be formed in supernova explosions. On the
terrestrial front, Fritz London was the first person to suggest that
superconductivity and superfluid flow in liquid helium were macroscopic
quantum phenomena. Then in 1950 Ginzburg and Landau proposed what seemed to be
a phenomenological theory by introducing a complex pseudo wave function $\psi$
as an order parameter in Landau's general theory of second-order phase
transitions. In 1957 the epoch-making microscopic Bardeen-Cooper-Schrieffer
(BCS) theory of superconductors was put forward, allowing
one to calculate the pairing gap $\Delta$. In it, the electron-phonon
interaction causes an instability in the Fermi-sea with respect to the
formation of what are now called {\it Cooper pairs}. Just
two years later, Gorkov showed that the Ginzburg-Landau theory was a limiting
form of the BCS theory around the critical temperature, in which $\psi$ is
directly proportional to the gap $\Delta$.

In the meantime, Bohr, Mottelson, and Pines had already (in 1958) 
made an analogy between the low-energy spectra of nuclei and
of the electrons in a superconducting metal. In 1959 Migdal,
\cite{Migdal:1959} in what was essentially a throw-away comment, noted that
the ``superfluidity of nuclear matter may lead to some interesting
cosmological phenomena if stars exist which have neutron cores. A star of this
type would be in a superfluid state with a transition temperature
corresponding to 1 MeV''. Thus, the study of
superfluidity in infinitely extended nuclear systems (such as neutron-star
matter or nuclear matter) predates the 1967 discovery of pulsars,
which were identified as rapidly rotating magnetic neutron
stars the next year. A few years after its formation, the
relevant part of a neutron star is expected to have a temperature from $10^6$
K to $10^9$ K, i.e. a small fraction of the pairing gap. 

Enter unconventional superfluidity: in 1972 Lee, Osheroff, and Richardson
 observed superfluid phases of $^3$He (a
fermionic atom) at a transition temperature of around 2.7 mK, three orders of
magnitude lower than for $^4$He. Just weeks after the discovery of the
superfluid, Anthony Leggett extended previous work of his
and identified the states seen in the new data. In a
similarly untypical development, in 1986 Bednorz and M\"{u}ller
 discovered superconductivity at 35 K in a barium lanthanum
copper oxide, initiating the era of what are now known as high-$T_c$
superconductors. Currently, the highest-$T_c$ superconductor is mercury
thallium barium calcium copper oxide at 138 K. In contradistinction to the
case of $^3$He, a comprehensive theory of high-$T_c$ superconductors is still
being sought.

In parallel developments, in 1980 Leggett showed that for
fermions the limits of tightly bound molecules and long-range Cooper pairs are
connected in a smooth crossover (BCS-BEC crossover). 
On the bosonic side, 1995 saw the first
realization of Bose-Einstein condensation in dilute vapors of alkali atoms. 
Most of the physics involved was
governed by mean-field interactions, and could be well described by the
Gross-Pitaevskii equation. As an experimental discovery it was very important
in and for itself, providing a clear realization of the phenomenon that
Einstein had theoretically predicted 70 years earlier. But the methods
involved in this research program on alkali atoms (magnetic trapping,
evaporative cooling, sympathetic cooling, optical trapping, and Feshbach
resonances) were then used to examine fermionic gases (in some cases, by the
same research groups), the first example of quantum degeneracy in trapped
Fermi gases being $^{40}$K in 1999 at JILA. In November
2003, three groups reported the realization of Bose-Einstein condensates of
molecules composed of fermionic atoms,
and soon after that the
observations were extended throughout the entire BEC-BCS crossover.
One of the high points (possibly {\it the} high
point) in the ensuing explosion of research was the demonstration of fermionic
superfluidity and phase coherence at MIT in April 2005. 
These phenomena, just like superfuidity in neutron matter, are characterized
by strong interactions and cannot be dealt with satisfactorily in a mean-field
theory framework. However, in contradistinction with neutron matter, cold 
atoms can be directly probed in the laboratory and can therefore help constrain
nuclear theory.

\begin{table}[b]
\caption{Typical values for the transition temperature $T_c$, and the 
ratio of the transition temperature to the Fermi temperature $T_c/T_F$, 
for various fermionic superfluids or superconductors.}
\begin{center}
\begin{tabular}{l r r}
\hline
 & $T_c$ & $T_c/T_F$ \\
\hline
Conventional superconductors &  5 K &$ $ $ $ $ $ 5 $\cdot 10^{-5}$\\
$^3$He  & 2.7 mK & 5 $\cdot 10^{-4}$ \\
High-$T_c$ superconductors  &  100 K & $10^{-2}$\\
Neutron matter &  $10^{10}$ K & 0.1\\ 
Atomic Fermi gases &$ $ $ $ $ $ 200 nK & 0.2\\
\hline
\end{tabular}
\end{center}
\label{table_Tc}
\end{table}

\section{Mise-en-sc\`{e}ne}
\label{section:INTRO_general}

The materials already referred to as high-$T_c$ superconductors are so
called because they have the highest $T_c$'s (in an absolute scale) currently
known on earth. However, if we take a different view and speak instead in
terms of relative temperatures, comparing the critical temperature with a
temperature that sets the scale in each system (the Fermi temperature $T_F$), things
look different. In Table \ref{table_Tc} we have summarized some results for
fermionic systems. Neutron matter and ultracold atomic gases are
separated by 17 orders of magnitude: however, in terms of their corresponding
Fermi temperatures, their critical temperatures are very large, essentially of
comparable magnitude. This is the reason both these systems can be viewed as
``high-temperature superfluids'', even though one occurs at very small and the
other at very large temperature. Since the pairing gap $\Delta$ 
is roughly proportional to the
critical temperature $T_c$, both these systems also exhibit very strong pairing, the
strongest ever observed before in nature or experimented with in the
laboratory. As we shall see below, in cold fermionic atoms the
particle-particle interactions can be tuned experimentally, allowing one to
examine a whole range of new phenomena, thus also mimicking the setting of
low-density neutron matter, which is beyond direct experimental reach.
In order to make this more tangible, let us begin with 
Schwinger and Bethe's ``shape-indendent formula'', used to 
describe $s$-wave scattering: 
\begin{equation}
k \cot \delta = -\frac{1}{a} + \frac{1}{2} k^2 r_e~,
\label{eq:shapeindependent}
\end{equation}
In this expression, $k$ is the momentum and $\delta$ is the $s$-wave phase shift. 
Knowing the latter gives us the scattering cross section, which is experimentally
observable. This formula implies that low-energy scattering provides information 
about only these two parameters: the scattering length $a$ and the 
effective range $r_e$. Thus, the cross section is independent of the details 
of the potential. Let us now discuss some specifics of our two systems and 
their interparticle interactions.

\subsection{Ultracold Atomic Gases}
\label{section:INTRO_general_CA}

For ultracold fermionic atoms
at low temperature, \cite{Ketterle:2008,Giorgini:2008,Grimm:2008,Luo:2009} 
the superfluid phase arises only in the
presence of interactions. 
The first batch of fermionic experiments was done with an equal mixture of the two
lowest hyperfine states, conventionally referred to as hyperfine states $| 1
\rangle$ and $| 2 \rangle$, of $^6$Li or $^{40}$K confined optically in a
dipole trap. 

The atom-atom interaction is quite complicated, though sophisticated models 
such as the Aziz potential for 
helium are often replaced by the simplified Lennard-Jones form. For cold 
alkali gases we have a van der Waals length $r_c \approx 50-100$ \AA, 
but in the quantum degenerate 
regime the interparticle spacing is at least a few thousand \AA, 
i.e. the interparticle spacing is significantly larger than the 
characteristic length of the interaction.
In a broad Feshbach resonance we have $k_F r_e \ll 1$, where $k_F$ is the Fermi wave vector
(connected to the number density via $\rho = k_F^3 / (3\pi^2)$), so at fixed density
the effective-range can be taken to be very small, essentially zero
(though it may be possible to use narrow and wide resonances in cold atoms to study
the case of varying $r_e$ experimentally, \cite{Marcelis:2008} and thus directly 
simulate neutron matter).

What was critically new in the fermionic cold atom experiments of the last
decade was the ability to vary the interaction strength (the scattering length $a$)
across a resonance,
through a regime known as ``unitarity''. This system is very interesting, since it is both
dilute (i.e. the range of the interatomic potential is much smaller than the
interparticle distance) and strongly interacting (i.e. the scattering length is
much larger than the interparticle distance). In this regime, all the length
scales associated with interactions drop out of the problem and the system is
expected to exhibit universal behavior, \cite{Bertsch:MBX} i.e. a
non-dependence on the specifics of the interatomic potential, which is why, in
principle, any fermionic atom can be used to perform such experiments. The
unitary Fermi gas has been the subject of intensive theoretical study, both in
the homogeneous case, and for the case of a
trapped gas. 

Let us now move on to observable many-body properties that can be and have
been measured in such experiments. Experiments using $^6$Li at
Duke University \cite{Luo:2009} and at ENS \cite{Navon:2010} have measured the ground-state
energy of the system, essentially finding it to be in agreement with Quantum
Monte Carlo predictions
\cite{Carlson:2003,Astrakharchik:2004,Carlson:2005,Gezerlis:2008,Forbes:2011} 
(very recent
experimental results from MIT \cite{Zwierlein:2011} and fermion-sign-free theoretical
values \cite{Carlson:2011} change 
this picture only quantitatively). 
The ground-state energy per particle is conventionally given in
units of the energy of a free Fermi gas at the same density $E_{FG} \equiv 3
E_F / 5 = 3 \hbar^2 k_F^2 / (10m)$ as $E = \xi E_{FG}$. 
Other experiments at MIT and Rice
probed lithium gases with population imbalance (also called ``polarized''
gases). An MIT
experiment \cite{Shin:2008} established the phase diagram of a polarized gas,
revealing spatial discontinuities in the spin polarization. This experiment
was then used \cite{Carlson:2008} to extract the pairing gap, which was found
to be approximately half of the Fermi energy $E_F$, in good agreement with QMC
calculations \cite{Carlson:2005}. (Similarly to the case of the ground-state
energy, the gap is conventionally given in units of $E_F$ as $\Delta = \eta
E_F$). The MIT group went on to use RF spectroscopy to independently determine
the gap, finding it to be in agreement with the afore-mentioned calculation
and extraction. \cite{Schirotzek:2008}

\subsection{Neutron-Star Matter}
\label{section:INTRO_general_NM}

The
neutrons in the inner crust are expected to pair in the $^1\mbox{S}_0$
channel, \cite{Lombardo:2001,Dean:2003}
but superfluidity in this channel will probably exhibit gap closure
before reaching the core. Let us note that even though, as we already pointed
out, the temperature of the crust is from $10^6$ K to $10^9$ K the neutron gas
is still superfluid because the critical temperature is expected to be larger,
approximately $10^{10}$ K (see Table \ref{table_Tc}). The first-order
approximation to this superfluid gas embedded in a lattice of nuclei is to
consider infinite pure neutron matter and examine the effect of the ion
lattice only at a later stage. For many decades, there has been a large spread
of predictions of the $^1\mbox{S}_0$ pairing gap even for this idealized pure
system, which we discuss in a later section. This spread is mainly due to 
the differences in the many-body schemes employed: dilute neutron matter
is a system that is very strongly paired, meaning that all perturbative
approaches fail, and most diagrammatic techniques in general have no
way of specifying why any specific class of diagrams should be dominant.

Even so, dilute neutron matter is simple when compared to the general
nuclear many-body problem. To see this, we turn 
turn our attention to the neutron-neutron interaction. 
In general it is complicated, necessitating
 sophisticated models (such as the Argonne $v18$ potential) to describe it. 
It has a strong dependence on quantum numbers such as total spin $S$ and isospin $T$. 
In local potentials it has a strongly repulsive core at distances of $\sim$ 0.5 fm, 
as well as an electromagnetic (in the case of neutrons, magnetic-moment) term, 
and one-pion exchange at large distances and intermediate-range spin-dependent 
attraction which is dominated by two-pion exchange.
However, as in the case of cold fermionic atoms, there are some simplifications:
in very dilute neutron matter the density (or $k_F$) goes to zero, making all
the terms in the interaction apart from the $s$-wave channel drop off. 
In contradistinction to the case of cold atoms, the neutron-neutron effective range
is not necessarily small in comparison to the interneutron spacing: $r_e = 2.7$ fm 
(with $k_F r_e$ going from 0 to 1.5). 
Another critical distinction is that for neutrons in a neutron-star crust the
scattering length is not at our disposal, being fixed, instead, at the value of
$a = -18.5$ fm, larger than the interparticle spacing ($k_F a$ goes from 0 to -10). 

Let us now move on to possible observational evidence of neutron-star 
matter pairing. Superfluidity in a neutron star is
often used to explain its dynamical and thermal evolution. Starting with the
former: the main example of this is given by pulsar post-glitch timing
observations. For some pulsars, even though their angular frequency decreases
slowly but steadily, there also occur sudden and rapid speed-ups of the
frequency (called glitches), which are in their turn followed by a gradual
decay of the increased frequency. The dominant phenomenological model holds
that the speed-up is triggered by a ``starquake'' in the crust. The crust
speed-up is rapidly communicated to the charged particles by the strong
magnetic field. However, the response of the neutron superfluid to the
speed-up is considerably slower and proceeds via electron scattering by the
normal fluid cores of the vortex lines in the rotating superfluid. (This model
was, astonishingly, proposed immediately after the first Vela pulsar glitch.)
\cite{Baym:1969}

The other commonly referred to aspect in which pairing is important has to do
with neutron-star cooling: observations of cooling quiescent neutron stars
provide evidence for the existence of a neutron superfluid throughout the
inner crust. \cite{Brown:2009} This is so because the specific heat of a
superfluid is exponentially suppresed. Furthermore, in the presence of a
neutron $^1\mbox{S}_0$ gap, the neutron-neutron bremsstrahlung reaction rate
is also suppressed. Finally, Cooper-pair breaking/formation neutrino emission
processes that occur near the transition temperature are also relevant to the
cooling of neutron stars, though due to the small size of the crust this
happens only during the first few hundred years of evolution, i.e. during the
crust's thermal relaxation. Cooper-pair breaking/formation calculations have
recently introduced vector current conservation, leading to the conclusion
that the vector part of the contribution to this process is strongly
suppressed, though the axial part is only slightly modified. These
three factors (specific heat, bremsstrahlung, pair breaking/formation) have
been incorporated in a recent minimal-cooling paradigm calculation,
\cite{Page:2009} which showed that different $^1\mbox{S}_0$ pairing gap
calculations lead to only slight quantitative differences in the cooling
curves of young neutron stars. All these factors show the importance of
superfluidity in neutron matter, but do not probe the magnitude of the pairing
gap. The situation in that
regard may be different as regards a new mechanism that makes use of
superfluid phonons. \cite{Aguilera:2009} Whether this mechanism is competitive
to the heat conduction by electrons in magnetized neutron stars or not is a
question that is directly correlated to the size of the gap.

Finally, let us point out that, outside the observational realm, neutron matter computations also hold 
significance in the context of traditional nuclear physics:
equation of state results at densities close to the nuclear saturation density
have been used for some time to constrain  Skyrme and other density functional approaches 
to heavy nuclei, \cite{Brown:2000} while the density-dependence 
of the $^1\mbox{S}_0$ gap in low-density neutron matter 
has recently also been used to constrain Skyrme-Hartree-Fock-Bogoliubov treatments 
in their description of neutron-rich nuclei \cite{Chamel:2008}.

\section{Quantum Many-Body Methods}

In the series of works reviewed in the present text, both cold atoms
and neutron matter were attacked using the same
methodology: Quantum Monte Carlo many-body simulations. 
Such methods have been used for decades to
calculate the energies of many systems, e.g. liquid helium. In such systems,
however, QMC methods were unable to provide the pairing gaps, because of the
vast difference in scale between the energy of the entire system (of the order
of eV's per particle) and the pairing gap (of the order of meV). Thus, in a
fortuitous turn of events, the same feature of strongly paired fermionic
systems (the large pairing gap) that precludes the application of mean-field
theories, is precisely the reason that allows many-body simulation techniques
to be used. Before giving a few technical details on how Green's Function
Monte Carlo works, we first go over the analytically exact expectations
for quantum many-body theory, that we know any correct new approach should
recover as a limiting case. After that we discuss BCS mean-field theory, 
which is not expected to be exact in the strongly paired regime, but even
so is a useful benchmark with which to compare other theories.

\subsection{Weak coupling}
\label{sec:exact} 

At extremely low densities ($| k_F a | << 1 $) the effective coupling between
two fermions is weak and matter properties can be calculated analytically.
The ground-state energy of normal (i.e. non-superfluid) matter in this regime
was calculated by Lee and Yang in 1957: \cite{Lee:1957}
\begin{equation}
\frac{E}{E_{FG}}  =  1 + \frac{10}{9\pi}k_Fa + \frac{4}{21\pi^2} \left
( 11 - 2 \ln2 \right ) \left ( k_Fa \right )^2~,
\label{eq:Lee}
\end{equation}
where $E_{FG}$ is the energy of a free Fermi gas at the same density as the
interacting gas.  While this expression ignores the contributions of
superfluidity, these are exponentially small in (1/$k_F a$).  In the next
section we compare these results to QMC calculations for $|k_F a| \geq 1$.

The pairing gap at weak coupling is also known analytically. The
mean-field BCS approach described below [Eq. \ref{deltaconteq}]
reduces in this limit to:
\begin{equation}
\Delta^0_{BCS}(k_F) = \frac{8}{e^2} \frac{\hbar^2 k_F^2}{2m} \exp\left( \frac{\pi}{2ak_F}\right)~.
\label{eq:weakBCS}
\end{equation}
However, as was shown in 1961 by Gorkov and
Melik-Barkhudarov, \cite{Gorkov:1961} the BCS result acquires a finite
polarization correction even at weak coupling, yielding a reduced pairing gap:
\begin{equation}
\Delta^0 (k_F) = \frac{1}{(4e)^{1/3}} \frac{8}{e^2} \frac{\hbar^2 k_F^2}{2m} \exp\left( \frac{\pi}{2ak_F}\right)~.
\label{eq:weakGMB}
\end{equation}
Thus, the polarization corrections reduce the mean-field BCS result by a factor of
$1 / (4e)^{1/3} \approx 0.45$. Interestingly, if one treats the polarization
effects at the level of sophistication used in the work of Gorkov and
Melik-Barkhudarov, this factor changes with $k_F a$ \cite{Schulze:2001},
though there is no {\it a priori} reason to expect such an approach to be
valid at stronger coupling ($k_F a$ of order 1 or more). Calculating the
pairing gap in this region has been an onerous task, as can be seen from the
multitude of publications devoted to this subject in the past few
decades.\cite{Lombardo:2001,Dean:2003,Chen:1993,Wambach:1993,
Schulze:1996,Schwenk:2003,Muether:2005,Fabrocini:2005,Cao:2006,
Margueron:2008,Gandolfi:2008,Abe:2009,Gandolfi:2009}

\subsection{BCS in the continuum and in a box}
\label{sec:bcs}
As the coupling strength increases, we expect the BCS mean-field theory to
become more accurate. In the BCS-BEC transition studied in cold atoms, the BCS
theory goes correctly to the two-body bound state equation in the deep BEC
regime. Though we do not expect BCS results to be quantitatively reliable 
in the regime of interest for neutron-star crusts, BCS theory
provides a standard basis of comparison for our {\it ab initio} results and
also allows us to analyze finite-size effects in the QMC simulations in a
simple way.  Within the BCS formalism the wave function is:
\begin{equation}
| \psi \rangle = \prod_{\bf k} (u_{{\bf k}} + v_{{\bf k}} c^{\dagger}_{{\bf k} \uparrow}
c^{\dagger}_{{-\bf k} \downarrow})|0\rangle~,
\label{eq:bcswave}
\end{equation}
where $u_{{\bf k}}^2 + v_{{\bf k}}^2 = 1$. A variational minimization of the
expectation value of the Hamiltonian for an average partice number (or
density) leads to the gap equation:
\begin{equation}
\Delta({\bf k}) = -\sum_{{\bf k'}} \langle {\bf k} |v | {\bf k'} \rangle \frac{\Delta({\bf k'})}{2E({\bf k'})}~,
\label{deltageneraleq}
\end{equation}
where the elementary quasi-particle excitations of the system have energy:
\begin{equation}
E({\bf k}) = \sqrt{\xi({\bf k})^2+\Delta({\bf k})^2}
\label{quasienereq}
\end{equation}
and $\xi({\bf k}) = \epsilon({\bf k})-\mu$, where the chemical potential is $\mu$ and $\epsilon({\bf k}) = \frac{\hbar^2k^2}{2m}$ 
is the single-particle energy of a particle with momentum ${\bf k}$. The
chemical potential is found by solving the gap equation together with the
equation that provides the average particle number:
\begin{equation}
\langle N \rangle = \sum_{{\bf k}} \left [ 1 - \frac{\xi({\bf k})}{E({\bf k})} \right ]~.
\label{particlegeneraleq}
\end{equation}
When interested in the $^1\mbox{S}_0$ gap for neutron matter, it is customary
to perform partial-wave expansions of the potential and the gap functions, as
well as an angle-average approximation. Thus, Eq. (\ref{deltageneraleq}) takes
the form:
\begin{equation}
\Delta(k) = -\frac{1}{\pi} \int\limits_0^{\infty} dk' ~k'^2 \frac{v(k,k')}{E(k')} \Delta(k')~,
\label{deltaconteq}
\end{equation}
where the potential matrix element is:
\begin{equation}
v(k,k') = \int\limits_0^{\infty} dr ~r^2 j_{0}(k'r) V(r) j_{0}(kr)~.
\end{equation}
Similarly, Eq. (\ref{particlegeneraleq}) becomes:
\begin{equation}
\rho = \frac{1}{2\pi^2} \int\limits_0^{\infty} dk ~k^2 \left(1 - \frac{\xi(k)}{E(k)}\right)~.
\label{particleconteq}
\end{equation}
These equations are one dimensional, and thus simple to treat numerically. The
density equation can be decoupled from the gap equation only when $\Delta /
\mu<<1$; this is not the case for the density regime we are considering. 

\begin{figure}[t]
\begin{center}
\includegraphics[width=0.8\textwidth]{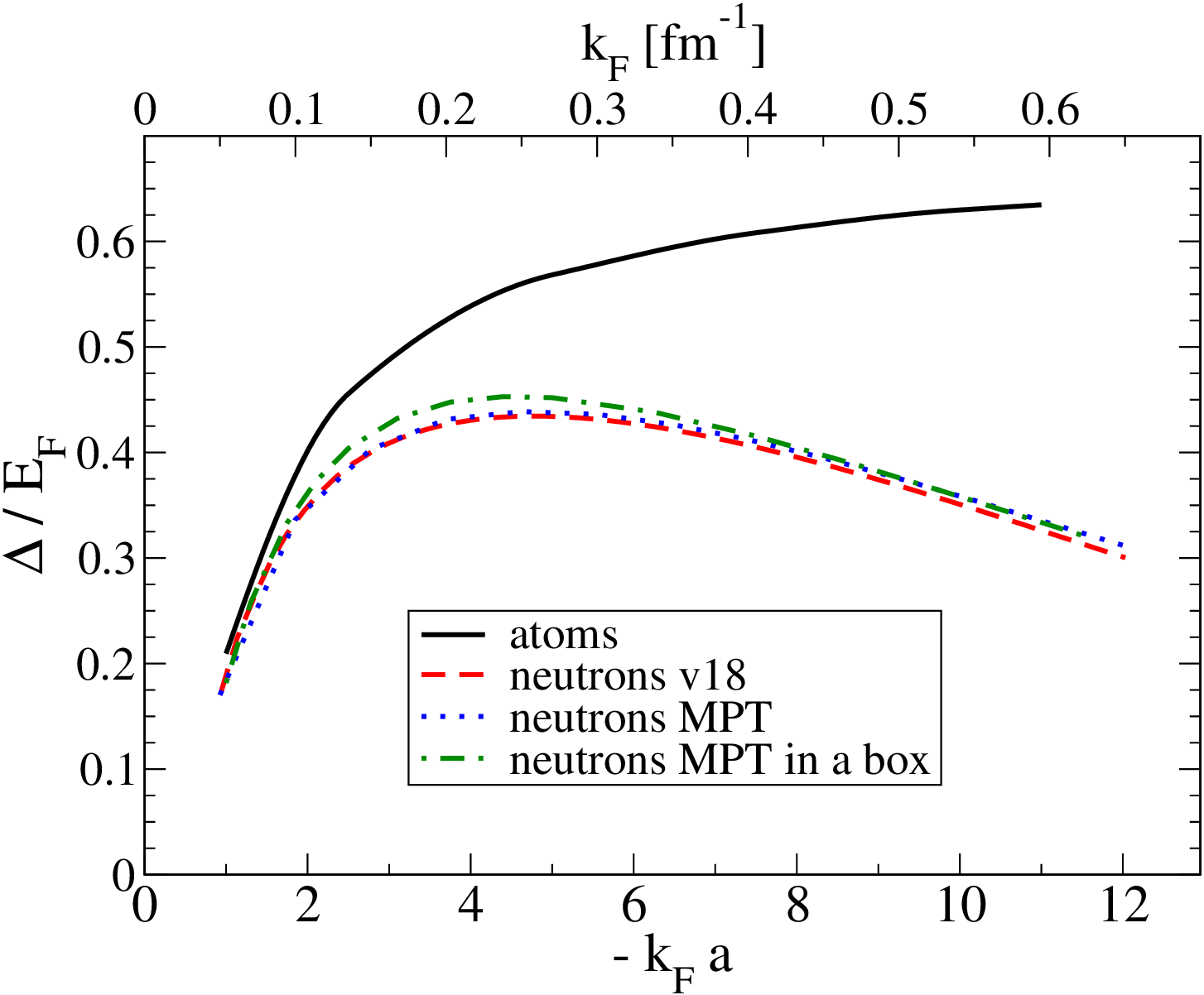}
\caption{BCS pairing gap $\Delta$ divided with
the Fermi energy $E_F$, versus the Fermi momentum $k_F$ for 
cold atoms (solid line), neutrons with AV18 (dashed line), neutrons
with a modified P\"{o}schl-Teller potential (dotted line) tuned to have the
same scattering length and effective range as AV18. At low density all
curves are identical for practical purposes. Also shown is the solution of the
BCS problem in a periodic box using the modified P\"{o}schl-Teller potential
for 66 neutrons (dash-dotted line).}
\label{fig:bcs}
\end{center}
\end{figure}

In Ref. \cite{Gezerlis:2008} Eq. (\ref{deltaconteq}) 
was solved in tandem with Eq.
(\ref{particleconteq}) for a modified P\"{o}schl-Teller (MPT) potential:
\begin{equation}
v(r) = - v_0 \frac{2 \hbar^2}{m} \frac{\nu^2}{\cosh^2(\nu r)}~,
\label{cosheq}
\end{equation}
where $v_0$ and $\nu$ are parameters which can be tuned so that this potential
reproduces any desirable scattering length and effective range. 
The potential in Eq. (\ref{cosheq})
clearly has no repulsive core, making it amenable to a straightforward
iterative solution. In this reference the potential was tuned so that the 
effective range is considerably shorter than the average interparticle
spacing and at the same time the product $k_F a$ was varied from -1 to -10.
The result is the solid line marked as ``atoms'' in Fig. \ref{fig:bcs}. If
$k_F a$ is taken to be infinite (the unitarity regime, unattainable
in neutron star crusts) the BCS limiting value turns out to be 0.69.

In Ref. \cite{Gezerlis:2010}  Eq. (\ref{deltaconteq}) was then solved 
together with Eq.
(\ref{particleconteq}) using the same MPT potential but the potential
parameters were tuned so as to reproduce the neutron-neutron 
scattering length $a \approx -18.5$ fm and effective range $r_e \approx 2.7$ fm,
the result also being shown in Fig. \ref{fig:bcs} (dotted line).
The same exercise was repeated for a more realistic microscopic potential,
namely the $^1\mbox{S}_0$ channel of the Argonne v18
\cite{Wiringa:1995} potential that contains a strong short-range repulsion.
This calculation is greatly simplified if one uses the method described in
Ref. \cite{Khodel:1996}, thereby transforming the problem into a quasilinear
one; the result is the dashed line.
For all the densities considered, the results of
solving Eqs. (\ref{deltaconteq}) and (\ref{particleconteq}) with the appropriately
tuned MPT potential and with the Argonne v18 potential are virtually indistinguishable.

The modified P\"{o}schl-Teller potential can also be used in a calculation for
finite average particle number. This is interesting because the Quantum Monte
Carlo simulations discussed below are performed in a periodic simulation volume.
This means that, in principle, there may exist considerable finite-size effects:
since the simulation cannot be performed in the thermodynamic limit, the question 
arises of how close to it one is. Refs. \cite{Gezerlis:2008, Gezerlis:2010, Gezerlis:2009}
solved Eqs. (\ref{deltageneraleq}) and
(\ref{particlegeneraleq}) for $\langle N \rangle$ from 20 to 200, in periodic
boundary conditions in a cubic box of volume $L^3$:
\begin{equation}
{\bf k}_{\bf n} = \frac{2 \pi}{L} ( n_{x}, n_{y}, n_{z})~.
\label{borneq}
\end{equation}
There it was
found that $\langle N \rangle = 66$ is very close to the thermodynamic limit.
In Fig.
\ref{fig:bcs} we show the results of solving the BCS gap equation Eq.
(\ref{deltageneraleq}) in a periodic box along with the particle-number
conserving Eq. (\ref{particlegeneraleq}) for $\langle N \rangle = 66$ (dash-dotted line);
as already noted, the discrepancy between that curve and the continuum result is small.
Given this finite-size analysis, the Quantum Monte Carlo results shown in the rest of
this review were performed using approximately 66 particles in periodic boundary conditions.

\subsection{Quantum Monte Carlo}
\label{sec:qmc}

The Hamiltonian for cold atoms and neutron matter at low densities is:
\begin{equation}
{\cal{H}} = \sum\limits_{k = 1}^{N}  ( - \frac{\hbar^2}{2m}\nabla_k^{2} )  + \sum\limits_{i<j'} v(r_{ij'})~.
\end{equation}
where $N$ is the total number of particles. In the case of cold atoms the interaction only
acts between opposite spin pairs and is of the form shown in Eq. (\ref{cosheq}).
The neutron-neutron interaction is more complicated but still simple if one considers
the AV4 formulation: \cite{Wiringa:2002}
\begin{equation}
v_4(r) = v_c(r) + v_{\sigma}(r){\mbox{\boldmath$\sigma$}}_1\cdot{\mbox{\boldmath$\sigma$}}_2,
\label{vfour}
\end{equation}
In the case of $S$=0 (singlet) pairs this gives:
\begin{equation}
v_S(r) = v_c(r) - 3v_{\sigma}(r)~.
\label{ves}
\end{equation}
However, it also implies an interaction for $S$=1 (triplet) pairs:
\begin{equation}
v_P(r) = v_c(r) + v_{\sigma}(r)~.
\end{equation}
Ref. \cite{Gezerlis:2010}  explicitly included such $p$-wave interactions in 
the same-spin pairs (the contribution of which was small even at the highest density considered), 
and perturbatively corrected the $S=1, M_S=0$
pairs to the correct $p$-wave interaction.

In these calculations it is customary to first employ a 
standard Variational Monte Carlo simulation. This approach
uses Monte
Carlo integration to minimize the expectation value of the Hamiltonian:
\begin{equation}
\langle H \rangle_{VMC} = \frac{\int d{\bf R} \Psi_{V}({\bf R}) H \Psi_{V}({\bf R})}{\int d{\bf R} |\Psi_{V}({\bf R})|^2} \geq E_0~.
\label{eq:qmcvmc}
\end{equation}
thereby optimizing a variational wave function $\Psi_V$.

At a second stage, the output of the Variational Monte Carlo calculation
is used as input in a fixed-node Green's Function Monte Carlo simulation,
which  
projects out the lowest-energy eigenstate
$\Psi_{0}$ from a trial (variational) wave function $\Psi_{V}$. This it does
by treating the Schr\"{o}dinger equation as a diffusion equation in imaginary
time $\tau$ and evolving the variational wave function up to large $\tau$.
The ground state is evaluated from:
\begin{eqnarray}
\Psi_0 & = & \exp [ - ( H - E_T ) \tau ] \Psi_V  \\ \nonumber
& = & \prod \exp [ - ( H - E_T ) \Delta \tau ] \Psi_V,
\end{eqnarray}
evaluated as a branching random walk.  The short-time propagator is usually
taken as
\begin{eqnarray}
\exp [ -H \Delta \tau ] = \exp [ -V \Delta \tau / 2 ] \exp [ -T \Delta \tau ] \exp [ -V \delta \tau / 2 ],
\label{eq:shorttime}
\end{eqnarray}
which is accurate to order $(\Delta \tau)^2$. 

The fixed-node calculation gives a wave function $\Psi_0$ that is the
lowest-energy state with the sames nodes (surface where $\Psi$ = 0) as the
trial state $\Psi_V$.  The resulting energy $E_0$ is an upper bound to the
true ground-state energy.  The variational wave function $\Psi_V$  has a
Jastrow-BCS form (see below), and contains a variety of parameters,
many of which affect the nodal surfaces. Since the fixed-node energy is an
upper bound to the true ground state, these parameters can be optimized to
give the best approximation to the ground-state wave function. In order to
optimize these variational parameters, they were included as slowly diffusing
coordinates in a preliminary GFMC calculation.  The parameters evolve slowly
in imaginary time, equilibrating around the lowest-energy state consistent
with the chosen form of the trial wave function.\cite{Carlson:2003}

The ground-state energy $E_0$ can be obtained from:
\begin{equation}
E_0 = \frac{ \langle \Psi_V | H | \Psi_0 \rangle}{ \langle \Psi_V | \Psi_0 \rangle}
= \frac{ \langle \Psi_0 | H | \Psi_0 \rangle}{ \langle \Psi_0 | \Psi_0 \rangle}.
\end{equation}

In all these Quantum Monte Carlo superfluid simulations the 
trial wave function was taken to be of the Jastrow-BCS form with fixed
particle number:
\begin{equation}
\Psi_V = \prod\limits_{i \neq j} f_P(r_{ij}) \prod\limits_{i' \neq j'} f_P(r_{i'j'}) \prod\limits_{i,j'} f(r_{ij'})  {\cal A} [ \prod_{i<j'} \phi (r_{ij'}) ]
\end{equation}
and periodic boundary conditions. The primed (unprimed) indices correspond to
spin-up (spin-down) neutrons. The pairing function $\phi (r)$ is a sum over
the momenta compatible with the periodic boundary conditions. In the BCS
theory the pairing function is:
\begin{equation}
\phi(r) =\sum\limits_{\bf n} \frac{v_{{\bf k}_{\bf n}}}{u_{{\bf k}_{\bf n}}} e^{i{\bf k}_{\bf n}\cdot{\bf r}} =\sum_{\bf n} \alpha_n e^{i{\bf k}_{\bf n}\cdot{\bf r}} ~,
\end{equation}
and can be parametrized with a short- and long-range part as in Ref.
\cite{Carlson:2003}:
\begin{equation}
\phi({\bf r}) = \tilde{\beta} (r) + \sum_{{\bf n},~I \leq I_C} \alpha_I e^{ i {\bf{k}}_{\bf n} \cdot {\bf r}}~,
\label{phieq}
\end{equation}
where $I = n_x^2 + n_y^2 + n_z^2$ using the parameters defined in Eq.
(\ref{borneq}). The Jastrow part is usually taken from a
lowest-order-constrained-variational method \cite{Pandharipande:1973}
calculation described by a Schr\"{o}dinger-like equation:
\begin{equation}
- \frac{\hbar^2}{m}\nabla^{2} f(r)  + v(r) f(r) = \lambda f(r)~\nonumber
\end{equation}
for the opposite-spin $f(r)$ and by the corresponding equation for the
same-spin $f_P(r)$. Since the $f(r)$ and $f_P(r)$ are nodeless, they do
not affect the final result apart from reducing the statistical error. The
fixed-node approximation guarantees that the result for one set of
pairing function parameters in Eq. (\ref{phieq}) will be an upper bound to the
true ground-state energy of the system. As mentioned above, 
the parameters are optimized in the full
QMC calculation, providing the best possible nodal surface, in the sense of
lowest fixed-node energy, with the given form of trial function.  
This upper-bound property is made use of in an attempt
 to get as close to the true ground-state energy as
possible.

\section{Equation of state}
\label{sec:eos}

\begin{figure}[t]
\begin{center}
\includegraphics[width=0.8\textwidth]{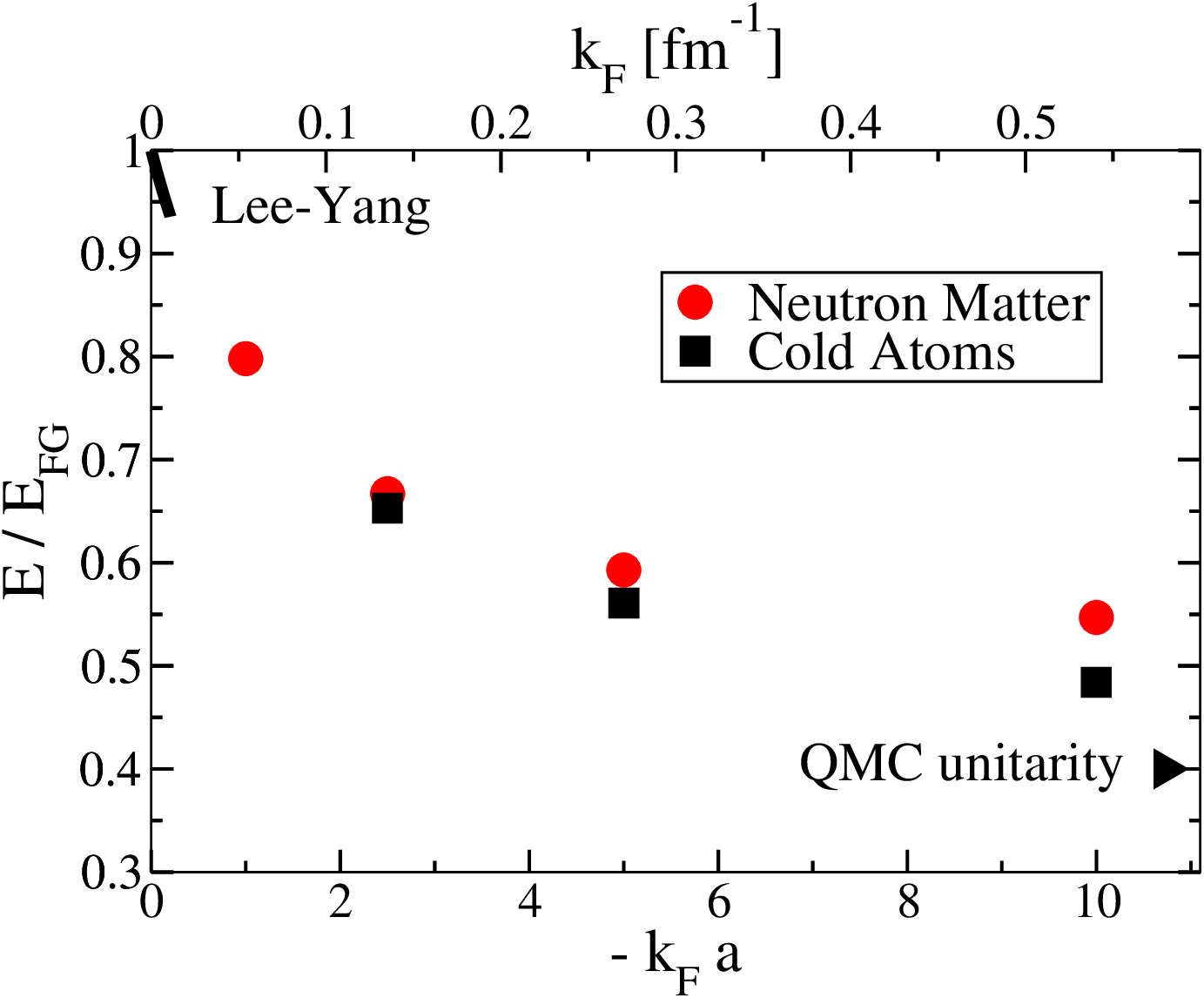}
\caption{Quantum Monte Carlo equation of state for cold atoms (squares) and neutron matter (circles). 
Also shown are the analytic expansion of the ground-state
energy of a normal fluid (line) and the Quantum Monte Carlo result at unitarity (arrow).}
\label{fig:eos_compare}
\end{center}
\end{figure}

The $T=0$ equations of state for cold atoms and neutron matter of Ref. \cite{Gezerlis:2008} are compared in Fig. \ref{fig:eos_compare}.
The horizontal axis is $k_F a$, with the equivalent Fermi momentum $k_F$ for neutron matter shown
along the top.  The vertical axis is the ratio of the ground-state energy to the free Fermi gas energy $(E_{FG})$ at the
same density; it must go to one at very low densities and decrease as the density increases and
the interactions become important.
The curve at lower densities shows the analytical result by Lee and Yang we mentioned in an earlier section. 
The QMC results shown in this figure seem to agree with the trend implied by the Lee-Yang result (which has not
been extended to even higher density because of its weak-coupling nature). 
The neutron matter and cold atom equations of state are very similar even for densities where the effective range is
comparable to the interparticle spacing. Hence, our earlier hope that cold-atom experiments can tell us something 
rather directly about the neutron-matter equation of state is now more than just a hope.
Near $k_F a = -10$ the energy per particle is not too far from QMC 
calculations
and measurements of the ratio $\xi$ of the unitary gas energy 
to $E_{FG}$ shown as an arrow on the right (it corresponds to $k_F a = \infty$); 
previous calculations give $\xi \approx 0.4$ (however, as mentioned above, see also Refs. \cite{Zwierlein:2011,Carlson:2011}).
At larger densities the cold-atom and neutron matter results start to diverge from each other.
This is due to two separate factors: i) the neutron finite effective range, and ii) 
the fact that the neutron results also incorporate an early attempt to correct the $S=1$, $M_S=0$ pairs.
When the density is very low, the $s$-wave
contribution is dominant so cold atoms and neutron matter agree very well.
At higher densities the energy is higher with the contribution of
the $p$-wave interaction.

\begin{figure}[t]
\begin{center}
\includegraphics[width=0.8\textwidth]{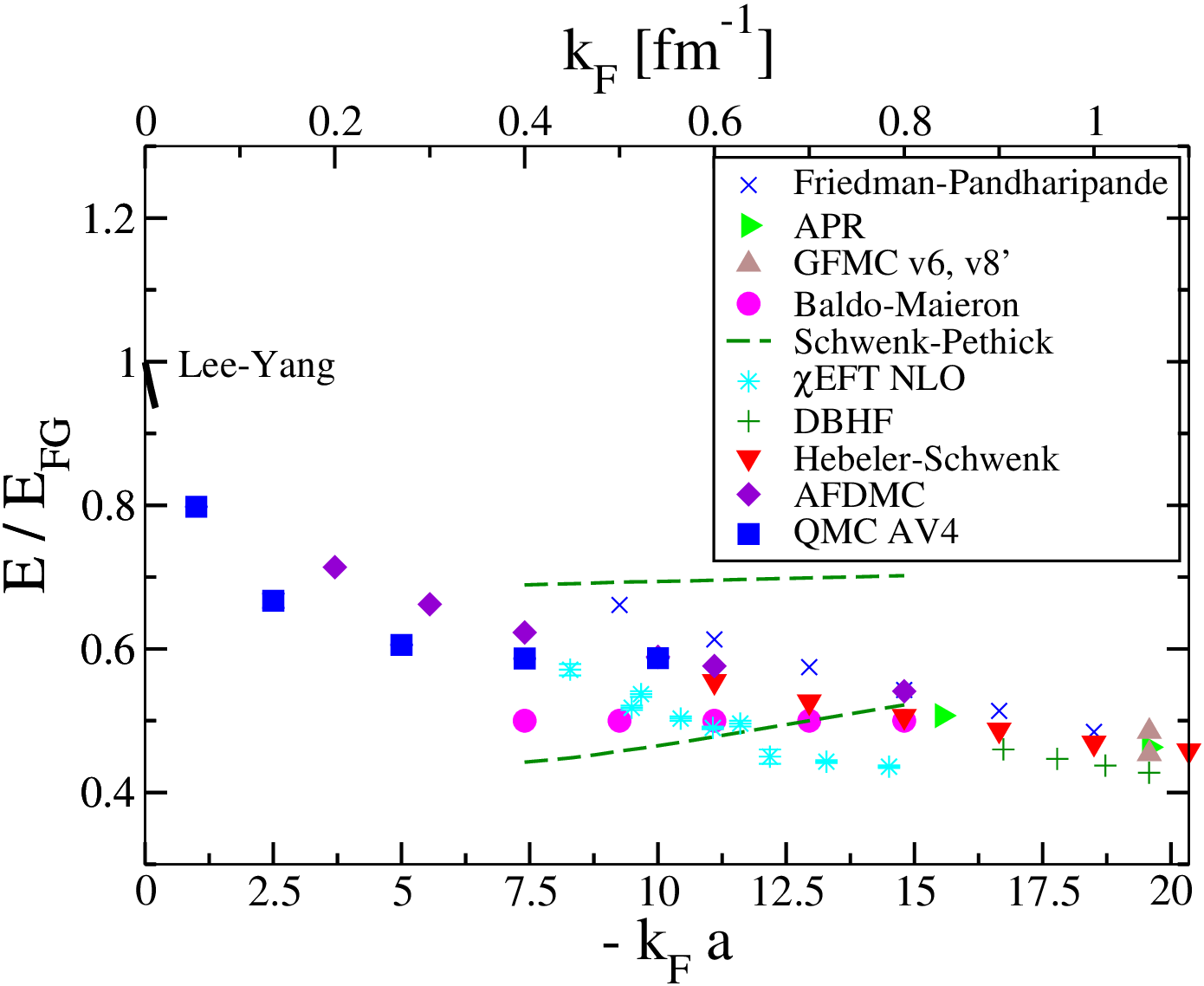}
\caption{Equation of state for neutron matter compared to
various previous results. Despite quantitative discrepancies, all calculations
give essentially similar results. Our lowest density corresponds to $k_F a =
-1$.}
\label{fig:DL}
\end{center}
\end{figure}

We also compare our neutron matter QMC AV4 results from Ref. \cite{Gezerlis:2010} for the
ground-state energy to other calculations extending to larger Fermi momenta.
Let us note that these are slightly different (larger) than the values shown in
Fig. \ref{fig:eos_compare} because of i) including same-spin $p$-wave contributions, and
ii) a more refined approach to dealing with the $S=1$, $M_S=0$ pairs.
For the highest density examined, $k_F a = -10$,
this change is approximately 7\%, while for $k_F a = -5$ it is  only 1\%.
Nonperturbative corrections at the highest density considered could reduce the
difference between the $s$-wave interaction and AV4 results slightly. 
In Fig. \ref{fig:DL} the results of Ref. \cite{Gezerlis:2010} are compared to (approximate) variational
hypernetted-chain calculations by Friedman and Pandharipande
\cite{Friedman:1981}, and another calculation by Akmal, Pandharipande, and
Ravenhall (APR) \cite{Akmal:1998}. Also included are two Green's Function Monte
Carlo results for 14 neutrons with more complete Hamiltonians
\cite{Carlson:Morales:2003}, a result following from a
Brueckner-Bethe-Goldstone expansion \cite{Baldo:2008}, a difermion EFT result
(shown are the error bands) \cite{Schwenk:2005}, the latest Auxiliary-Field
Diffusion Monte Carlo (AFDMC) result (discussed below) \cite{Gandolfi:2008}, a
Dirac-Brueckner-Hartree-Fock calculation \cite{Margueron:2007b}, a lattice
chiral EFT method at next to leading order \cite{Epelbaum:2008b} (see also
Ref. \cite{Epelbaum:2008a}), and an approach that makes use of chiral N$^2$LO
three-nucleon forces.\cite{Hebeler:2010} Of these, Refs. \cite{Akmal:1998},
\cite{Gandolfi:2008}, and \cite{Hebeler:2010} include a three-nucleon
interaction, though at the densities considered, these are not expected to be
significant.  Qualitatively all of these results agree within 20\%.  

\section{Pairing gap}
\label{sec:gap}

\begin{figure}[t]
\begin{center}
\includegraphics[width=0.8\textwidth]{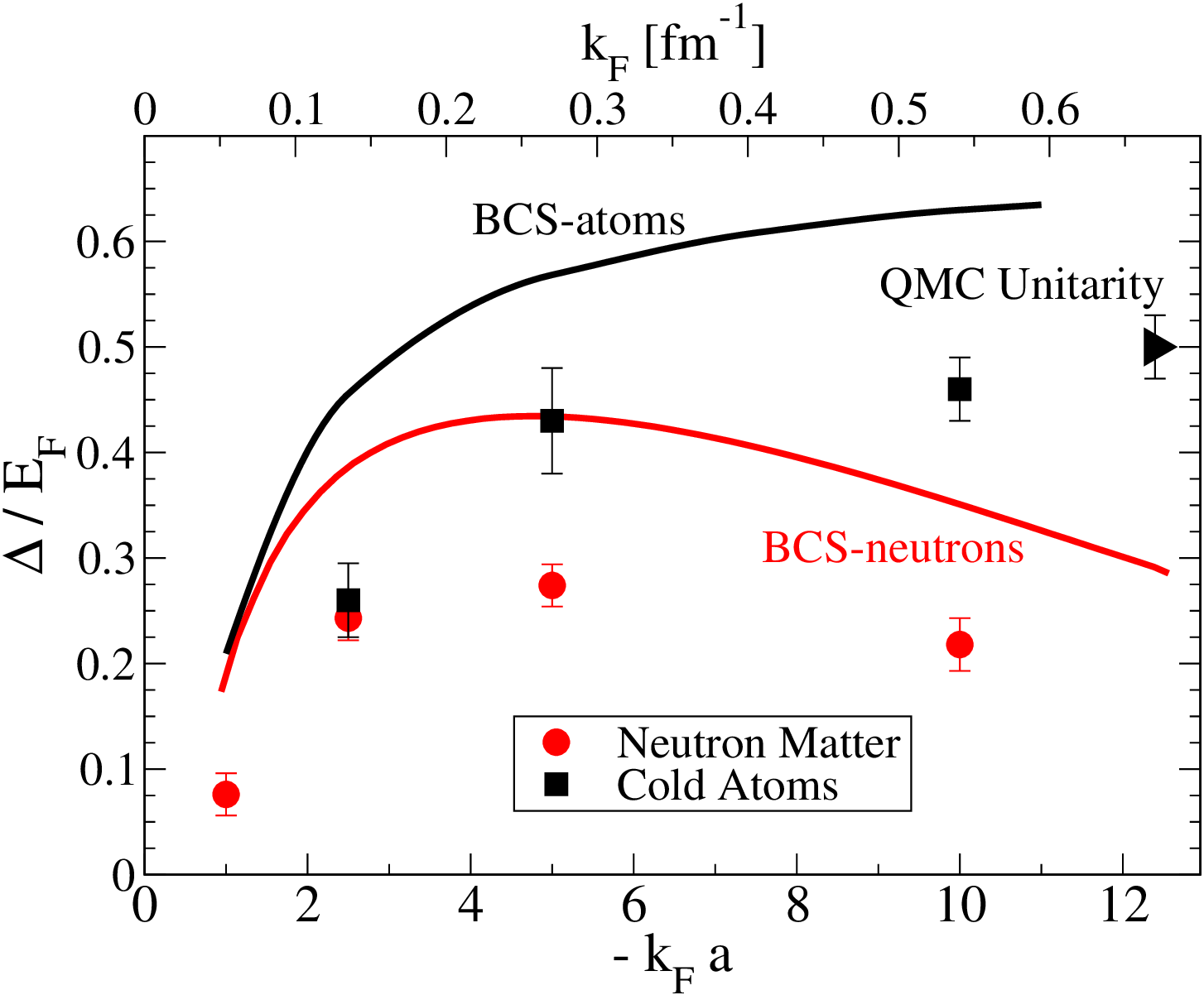}
\caption{Superfluid pairing gaps versus $k_F a$ for cold atoms ($r_e \approx 0$) 
and neutron matter ($ |r_e / a| \approx 0.15$).
BCS (solid lines) and QMC results (points) are shown.}
\label{fig:AFDMClikegap}
\end{center}
\end{figure}

In Fig. \ref{fig:AFDMClikegap} we plot the pairing gap as a function of $k_F a$
for both cold atoms and neutron matter, taken from Ref. \cite{Gezerlis:2008}. 
The Quantum Monte Carlo pairing gaps are arrived at from the ground-state
energy, through the use of the odd-even staggering formula:
\begin{equation}
\Delta = E(N+1) - \frac{1}{2} \left [ E(N)+E(N+2) \right ]~,
\label{eq:staggerer}
\end{equation}
where $N$ is an even number of particles. 
BCS calculations are shown as solid
lines, and QMC results are shown as points with error bars. QMC pairing gaps
are shown from calculations of $N=66-68$ particles.
For very weak coupling,
$ - k_F a << 1$, the pairing gap is expected to be reduced from the BCS
value by the polarization corrections calculated by Gorkov and
Melik-Barkhudarov, $\Delta/ \Delta_{BCS} = (1/4e)^{1/3}$, 
as mentioned in an earlier section (Eq. (\ref{eq:weakGMB})). 
Because of finite-size effects, it is difficult
to calculate pairing gaps using QMC in the weak coupling regime.
The QMC calculations at the lowest density, $k_F a = -1$, are roughly
consistent with this reduction from the BCS value.
At slightly larger yet still small densities,
where $-k_F a = {\cal O} (1)$ but $k_F r_e << 1$ for neutron matter, 
one would expect the pairing gap to be similar for cold atoms and neutron 
matter.  The results at $k_F a = -2.5$, where $k_F r_e \approx 0.35$, support 
this expectation. Beyond that density the effective range becomes important 
and the QMC results are significantly reduced in relation to the cold atoms where $r_e \approx 0$.

In cold atoms, the suppression from BCS is reduced as the density increases, with a
smoothly growing fraction of the BCS results as we move from the BCS to the
BEC regime.  At unitarity the measured pairing gaps
\cite{Shin:2008,Carlson:2008,Schirotzek:2008} are 0.45(0.05) of the Fermi
energy, for a ratio $\Delta/\Delta_{BCS} \approx 0.65$,  in agreement with
predictions by QMC methods.\cite{Carlson:2003,Carlson:2005,Gezerlis:2008} In
neutron matter, though, the finite range of the potential reduces $\Delta/E_F$
as the density increases.  We find a ratio $\Delta/\Delta_{BCS}$ that
increases slightly from $|k_F a| = 1$ to $2.5$, but thereafter remains roughly
constant.

\begin{figure}[t]
\begin{center}
\includegraphics[width=0.8\textwidth]{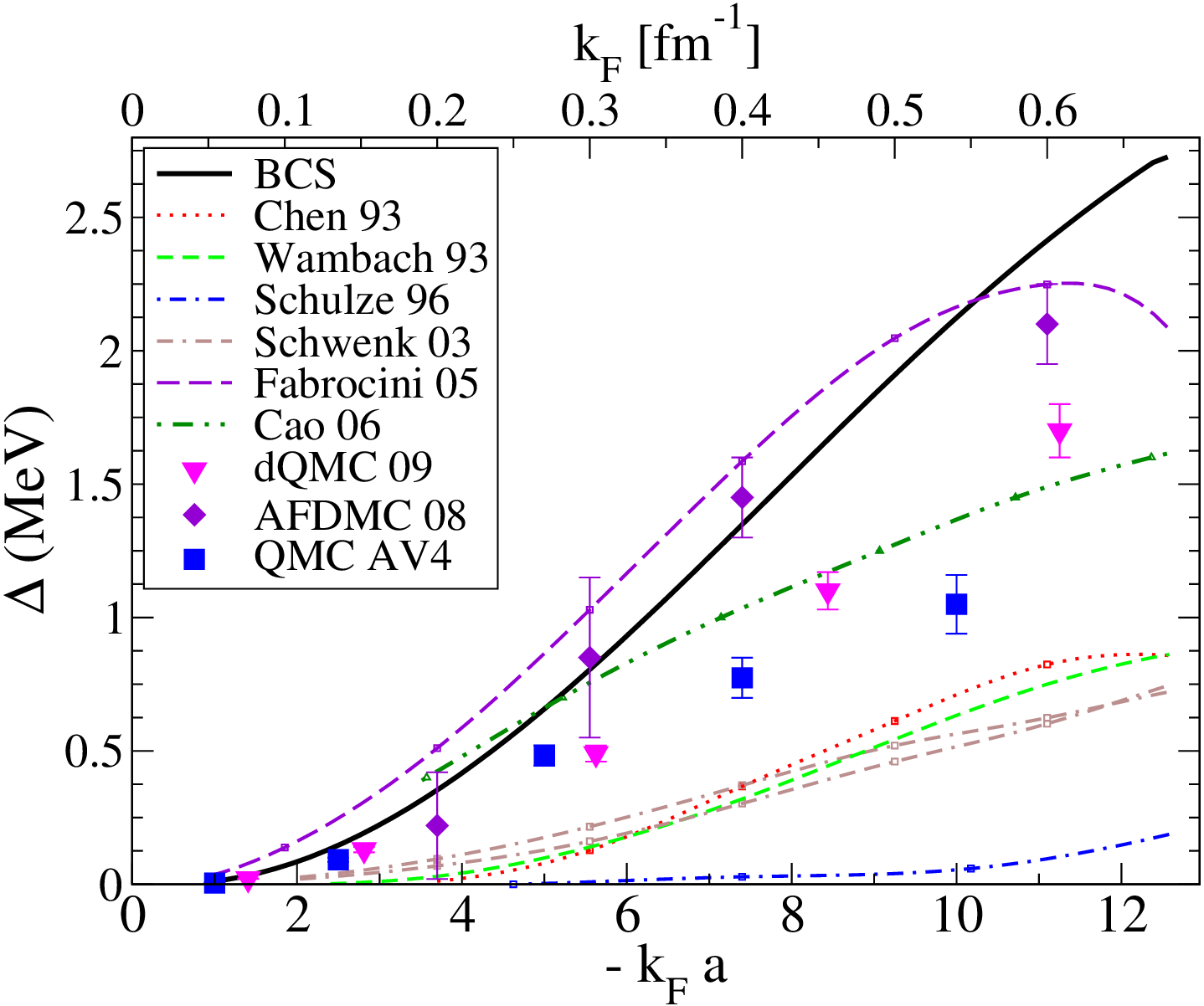}
\caption{Superfluid pairing gap versus $k_F a$ for neutron
matter compared to previous results.}
\label{fig:gap-compare}
\end{center}
\end{figure}

In Fig. \ref{fig:gap-compare} we compare the neutron matter results of Ref. \cite{Gezerlis:2010} 
to selected previous
results: a Correlated-Basis Function calculation by Chen {\it et al.}
\cite{Chen:1993}, an extension of the polarization-potential model by Wambach
{\it et al.} \cite{Wambach:1993}, a medium-polarization calculation by Schulze
{\it et al.} \cite{Schulze:1996}, a renormalization group calculation by
Schwenk {\it et al.} \cite{Schwenk:2003}, a Brueckner calculation by Cao {\it
et al.} \cite{Cao:2006}, a determinantal lattice QMC approach \cite{Abe:2009},
and finally the newer Correlated-Basis Function calculation by Fabrocini {\it
et al.} \cite{Fabrocini:2005} that was used as an input wave function in the
two AFDMC calculations of 2005 and 2008.\cite{Fabrocini:2005,Gandolfi:2008} 

The results of our QMC calculations are much larger than most diagrammatic
\cite{Chen:1993,Wambach:1993,Schulze:1996} and renormalization group
\cite{Schwenk:2003} approaches. As these approaches assume a well-defined
Fermi surface or calculate polarization corrections based on single-particle
excitations it is not clear how well they can describe neutron matter in the
strongly paired regime, or the similar pairing found in cold atoms. Ref.
\cite{Cao:2006}, which incorporates self-energy corrections and screening at
the RPA level within Brueckner theory, appears to give results similar to
ours. However, these values disagree with our lower-density results and,
perhaps more importantly, at the lowest density reported the gap is larger
than the mean-field BCS value (which is disallowed, as discussed in
 section \ref{sec:exact}). On a similar
note, Refs. \cite{Wambach:1993} and \cite{Schwenk:2003} make use of a
weak-coupling formula to calculate the pairing gap, similarly to Eqs.
(\ref{eq:weakBCS}) and (\ref{eq:weakGMB}). 
The prefactor they use is justified based on predictions in
the theory of $^3$He. However, the concept itself of a Fermi surface is not
well-defined in these strongly paired systems: in $^3$He, in contrast to the
present case, the gap is considerably smaller than the Fermi energy.

Having discussed the relationship of our QMC results to analytical approaches,
a few comments on the relationship between the  different extant simulation results are in order.
The first AFDMC results for pairing gaps in neutron
matter were published in Ref. \cite{Fabrocini:2005}. These initial calculations
were for very small system sizes of N=14-20 and yielded quite
large pairing gaps. The AFDMC
group then repeated their calculations for larger
systems, \cite{Gandolfi:2008} reducing the gap and eliminating part of the difference
with GFMC results, but not entirely. Detailed explanations
are given in Ref. \cite{Gezerlis:2010}. The AFDMC approach includes a
realistic two plus three nucleon interaction, and is therefore valuable for moderate
to large densities. At very small densities we expect the s- and p-wave
interactions to be sufficient. The trial wave function (and the constrained path approximation) for the
AFDMC method are chosen from another approach. In the
calculations of Refs. \cite{Fabrocini:2005,Gandolfi:2008} the
wave function was taken from a Correlated-Basis Function (CBF) approach.
This wave function is unrealistic at very low density as it yields a pairing
gap larger than the Fermi energy.
The  QMC AV4 results use a wave function that has been variationally
optimized, but a much simpler Hamiltonian that should nevertheless
be valid at the lowest densities.
Ref. \cite{Gezerlis:2010} demonstrates that QMC calculations
of the gap can be  quite sensitive to the trial function,
for example using the CBF input wave function the QMC calculation
reproduces the AFDMC result for the pairing gap. We expect the
QMC results are more accurate at low densities, but at higher densities
the full interaction treated with AFDMC is required, along with a careful study of the impact of the trial wave function.

Finally, the QMC results seem to qualitatively agree (at least for the lowest densities
considered) with a determinantal Quantum Monte Carlo lattice calculation
\cite{Abe:2009} which, however, only includes the $s$-wave component in the
interaction (which is why they agree much better with 
the results of Ref. \cite{Gezerlis:2008} than those of Ref. \cite{Gezerlis:2010}). 
This implies that a consensus is emerging, in that both the GFMC and the dQMC
approaches find a gap that is suppressed with respect to the mean-field BCS
result but is still a substantial fraction of the Fermi energy.

\section{Conclusions and Future Directions}
\label{sec:future}

To conclude, we have calculated the equation of state and pairing gap of cold atoms and
low-density neutron matter with the AV4 interaction from $|k_F a|$ from $1$ to $10$. 
The calculated equation of state and pairing gap match smoothly with the known 
analytic results at low densities, and provide important constraints in 
the strong-coupling regime at large $k_F a$.  
The low-density equation of state can help constrain 
Skyrme mean-field models of finite nuclei. The pairing gap for 
low-density neutron matter is relevant to Skyrme-Hartree-Fock-Bogoliubov 
calculations \cite{Chamel:2008} of neutron-rich nuclei and to 
neutron-star physics, since it is expected to influence the 
behavior of the crust. \cite{Page:2009}
Moreover, a magnetic field in the neutron star crust would 
have to be approximately $10^{17}$ G to overcome this gap and thus 
polarize neutron matter; such a value of the magnetic field is not 
implausible within the context of magnetars.
The question of polarized neutron matter also has a long history \cite{Haensel:1975}
and has recently been attacked using Quantum Monte Carlo. \cite{Gezerlis:2011,Gezerlis:2012}
Similarly, the fact that the magnitude of the gap is not as small 
as previously expected implies that a new mechanism that makes 
use of superfluid phonons is competitive to the heat conduction 
by electrons in magnetized neutron stars.\cite{Aguilera:2009}  

This line of Quantum Monte Carlo calculations, having first been applied to and verified in cold atomic experiments, can
also provide directions for future work in the field of nucleonic infinite matter. 
The simplest case is that of a two-component gas where the
two populations are equal.\cite{Gezerlis:2008,Gezerlis:2010} The next step is to examine the ramifications of taking
different populations for the two components: this is the case of spin-polarized low-density neutrons studied in Refs. \cite{Gezerlis:2011,Gezerlis:2012}.
Cold-atom experiments have by now also addressed Efimov physics, in which three components are involved. In the nuclear context,
adding a third species could provide further insight into the physics of neutron stars. If the third component particles
were taken to be protons and, as in this paper, only a few of them were added, then it would be possible to study 
highly asymmetric nuclear matter. Another possible avenue of future research is related to optical lattice experiments with cold atoms:
to first approximation these are equivalent to periodic external potentials. In the nuclear case, an external potential
would allow us to study the static response of neutron matter and would also facilitate the understanding of the impact
on neutron pairing of the ion lattice that exists in a neutron star crust.
Such microscopic results for the static response could provide further constraints on energy-density functionals used
to describe the crust of neutron stars.

\vspace{0.5cm}

{\bf Acknowledgments} 
The authors have benefited
from a series of conversations on these topics with Stefano Gandolfi,
Sanjay Reddy, and Achim Schwenk. This work was supported by 
DOE Grant Nos. DE-FG02-97ER41014 and DE-AC52-06NA25396.
The computations shown were performed at the National Energy
Research Scientific Computing Center (NERSC) and through
Los Alamos Supercomputing.


\label{lastpage-01}

\end{document}